\documentclass[%
reprint,
 amsmath,amssymb,
 aps,
prl,
showkeys
]{revtex4-2}

\usepackage{booktabs}
\usepackage{graphicx}
\usepackage{hyperref}
\hypersetup{
  colorlinks,%
  citecolor=black,%
  filecolor=black,%
  linkcolor=black,%
  urlcolor=blue
  }
\usepackage{float}
\usepackage{multirow}
\usepackage{xspace}

\usepackage{bm}

\newcommand*{\sqs}{\ensuremath{\sqrt{s}}\xspace}

\newcommand*{\Nch}{\ensuremath{N_\mathrm{ch}}\xspace}

\newcommand*{\SO}{\ensuremath{S^{p_{\rm T}=1}_\mathrm{0}}\xspace}
\newcommand*{\pT}{\ensuremath{p_\mathrm{T}}\xspace}
\newcommand*{\mT}{\ensuremath{m_\mathrm{T}}\xspace}

\newcommand*{\meanpT}{\ensuremath{\left<\pT\right>}\xspace}

\newcommand*{\dNch}{\ensuremath{{\rm d}N_{\rm ch}/{\rm d}\eta}\xspace}
\newcommand*{\dd}{\ensuremath{\mathrm d}\xspace}

\usepackage{orcidlink}

\newcommand{\orcidA}{\orcidlink{0000-0002-2420-7650}} % Add \orcidA{} behind the author's name // LGy
\newcommand{\orcidB}{\orcidlink{0000-0003-2849-0120}} % Add \orcidB{} behind the author's name // GB
\newcommand{\orcidC}{\orcidlink{0000-0001-9223-6480}} % Add \orcidC{} behind the author's name // GGB
\newcommand{\orcidD}{\orcidlink{0000-0003-3706-5265}} % Add \orcidD{} behind the author's name // RV

\begin{document}

\preprint{APS/123-QED}

\title{Effect of event classification on the Tsallis-thermometer}

\author{L\'aszl\'o Gyulai\orcidA{}}
\author{G\'abor B\'ir\'o\orcidB{}}
\altaffiliation[]{ELTE E\"otv\"os Lor\'and University, Institute of Physics, 1/A P\'azm\'any P\'eter S\'et\'any, H-1117  Budapest, Hungary.}
\author{R\'obert V\'ertesi\orcidD{}}
\email{vertesi.robert@wigner.hun-ren.hu}
\author{Gergely G\'abor Barnaf\"oldi\orcidC{}}
\affiliation{HUN-REN Wigner Research Center for Physics, 29--33 Konkoly--Thege Mikl\'os Str., H-1121 Budapest, Hungary}

\date{\today}

\begin{abstract}
We analyze identified hadron spectra in pp collisions at $\sqs = 13$~TeV measured by ALICE within a non-extensive statistical framework. Spectra classified by multiplicity, flattenicity, and spherocity were fitted with the Tsallis–-Pareto distribution, and the parameters were studied on the Tsallis-thermometer. Multiplicity and flattenicity classes follow a previously observed scaling, while the non-extensivity parameter shows a distinct sensitivity to the spherocity. A data-driven parametrization confirms a proportionality between the Tsallis temperature and mean transverse momentum, offering a simple estimate of the effective temperature. These results highlight the ability of the Tsallis-thermometer to capture both multiplicity and event-shape effects, linking soft and hard processes in small systems.
\end{abstract}

\maketitle

\section{Introduction}

In high-energy heavy-ion collisions at the Relativistic Heavy Ion Collider (RHIC) and Large Hadron Collider (LHC), a new state of matter can be created where the quarks are no longer confined into hadrons. This extremely hot and dense medium, dubbed the quark--gluon plasma (QGP), is found to exhibit fluid-like collective motion~\cite{PHENIX:2004vcz,STAR:2005gfr,PHENIX:2008uif,ALICE:2022wpn}. In recent years, similar collective behavior has been found in small collision systems such as proton--proton and proton--nucleus, with high final-state multiplicity~\cite{CMS:2010ifv,ATLAS:2012cix,ALICE:2012eyl}. This poses the question whether QGP can be created in small systems. Several approaches have been developed to resolve this question, based on hydrodynamical description~\cite{Bozek:2011if,Nagle:2018nvi,Romatschke:2016hle}, color glass condensate explanations~\cite{Dusling:2013oia}, and vacuum-like QCD effects~\cite{OrtizVelasquez:2013ofg,Huo:2017nms}. 

Transverse-momentum distributions (spectra) of identified hadrons comprise a thermal-like soft part and a power-law-like regime that is attributed to hard perturbative chromodynamics (pQCD) processes. These two parts can be described in a unified framework, from small to large collision systems, using non-extensive thermodynamics~\cite{Tsallis:1987eu, Tsallis:2009tri}. 
Long-standing investigations into high-energy hadron spectra within the framework of non-extensive Tsallis statistics span both theoretical developments and phenomenological fits across collision systems and energies~\cite{Bhattacharyya:2017hdc, Biro:2020kve, Patra:2020gzw}. 
The Tsallis-thermometer defines the effective temperature of a system within the framework of non-extensive thermodynamics, where particle distributions follow a Tsallis--Pareto form characterized by two parameters: the temperature $T$ and the non-extensivity parameter $q$. The $q\rightarrow1$ case returns the Boltzmann--Gibbs distribution and its associated Boltzmann temperature~\cite{Biro:2020kve}. Using the Tsallis-thermometer, one can gain insight into several properties of the system, such as the system size, the timeline of the formation of different particle spectra, and heat capacity~\cite{Gyulai:2024qov,Gyulai:2024dkq}. Moreover, the mean transverse momentum, when compared across collision systems of different sizes, offers a direct test of the scaling properties predicted within the Tsallis--Pareto non-extensive framework, and reveals how system size influences non-extensivity and effective temperature~\cite{ALICE:2022xip, Biro:2024syz, Gardim:2024zvi}.

In high-energy physics, event classification provides means to group events by global properties such as multiplicity, energy density, or centrality. These classes can be used to select events with specific thermodynamic properties such as temperature, entropy, or phase space, allowing statistical and thermodynamic models to describe the system’s evolution. 
Event classification is a useful tool to distinguish between QGP and vacuum-QCD scenarios~\cite{OrtizVelasquez:2013ofg,Nagle:2018nvi}. Event shape variables become especially interesting in understanding data from recent light-ion collisions such as O--O or Ne--Ne, which may probe the onset of QGP effects~\cite{CMS:2025tga,CMS:2025bta,ALICE:2025luc,ATLAS:2025nnt}.

However, non-extensive parameters extracted from minimum-bias spectra generally average over a mixture of event types, potentially blending together contributions from different microscopic environments. This raises the question, to what extent do the Tsallis parameters encode genuine non-equilibrium features of the particle-emitting source, and to what extent are they shaped by the mixture of event classes included in the sample. Event classification therefore provides a natural way to refine the interpretation of the Tsallis parameters by constructing narrower, more homogeneous ensembles. Studying how the extracted $T$ and $q$ values evolve across event classes with different global properties can help disentangle the contributions of system size, semi-hard activity, and possible collective effects---as summarized in review~\cite{Prasad:2025yfj}.

In the current work, we use the non-extensive thermodynamical framework together with event classification to obtain insights into the processes that shape the particle spectra.

\section{Event classification}
\label{sec:classify}

One of the simplest event classifiers is event multiplicity, the number of final state particles that can be taken at central or forward rapidity. 
It has been noticed that several observables primarily scale with multiplicity (\Nch) across several collision systems, such as light and strange particle ratios and angular correlations~\cite{ALICE:2016fzo,ALICE:2019zfl}. Multiplicity from the forward region correlates strongly with the event multiplicity, but is less influenced by hard processes at midrapidity. To decrease autocorrelation effects, ALICE categorizes events by multiplicity measured by the V0 detectors~\cite{ALICE:2013axi} in the forward regions, denoted by V0M. Note that with Run 3, V0 has been retired and the Fast Interaction Trigger (FIT) system is used for event characterization instead~\cite{Roslon:2025lxb}.

Transverse spherocity is an event-shape observable that characterizes the geometry of particle momentum distributions in the transverse plane. 
It has been widely used in pp and p–Pb collisions to separate hard-scattering-dominated events from soft, isotropic ones~\cite{Ortiz:2015ttf}.
In the current work we use the definition of the unweighted transverse spherocity,  
\begin{equation}
    \SO = \frac{\pi^{2}}{4} \; \min_{\hat{n}_{\rm T}} \left( \frac{\sum_{i=1}^{N_{\rm trk}} \left|\hat{p}_{{\rm T},i} \times \hat{n}_{\rm T} \right|}{N_{\rm trk}} \right)^{2},
\end{equation}
where $\hat{p}_{{\rm T},i}$ is the unit vector in the direction of the transverse momentum of particle $i$, $\hat{n}_{\rm T}$ is the unit vector in the transverse plane of a given event that minimizes the normalized sum of transverse momentum projections perpendicular to it, and $N_{\rm trk}$ is the number of tracks. This definition ensures that \SO is a purely geometric quantity that takes each track into account with equal weight. Transverse spherocity ranges from 0 for highly jet-like events with collimated momentum flow to 1 for perfectly isotropic events.

Flattenicity is a novel event-shape observable that quantifies how evenly charged-particle activity is distributed in the pseudorapidity ($\eta$) -- azimuth ($\varphi$) space. ALICE uses a definition of flattenicity where the $\eta$--$\varphi$ plane is divided into 1440 cells of equal size (40 divisions in $\eta$ and 36 in $\varphi$): 
\begin{equation}
\rho = \frac{ \sigma^{2}_{N_{\rm cell}} }{ \left< N_{\rm cell} \right>^{2} },
\end{equation}
where $\left< N_{\rm cell} \right>$ is the average and $\sigma^{2}_{N_{\rm cell}}$ is the variation of cell multiplicities.
Lower flattenicity values correspond to more uniform (flat) distributions, while non-uniform (spiky) events have high flattenicity~\cite{Ortiz:2022mfv}. 
To better associate flattenicity with other event shape variables, $1-\rho$ is often utilized instead of $\rho$ in practical applications~\cite{ALICE:2024vaf}. We will also use this notation in the followings.

In the current work, we examine the behavior of identified hadrons on the Tsallis-thermometer in terms of various event classification methods.
We use charged-hadron multiplicity in the forward range, transverse spherocity, and flattenicity to classify events.

To explore the connection between the event classifier variables, we used PYTHIA~8 to simulate the behavior of flattenicity and spherocity in correlation with the event multiplicity.
\begin{figure}[h!]
\includegraphics[width=0.48\textwidth]{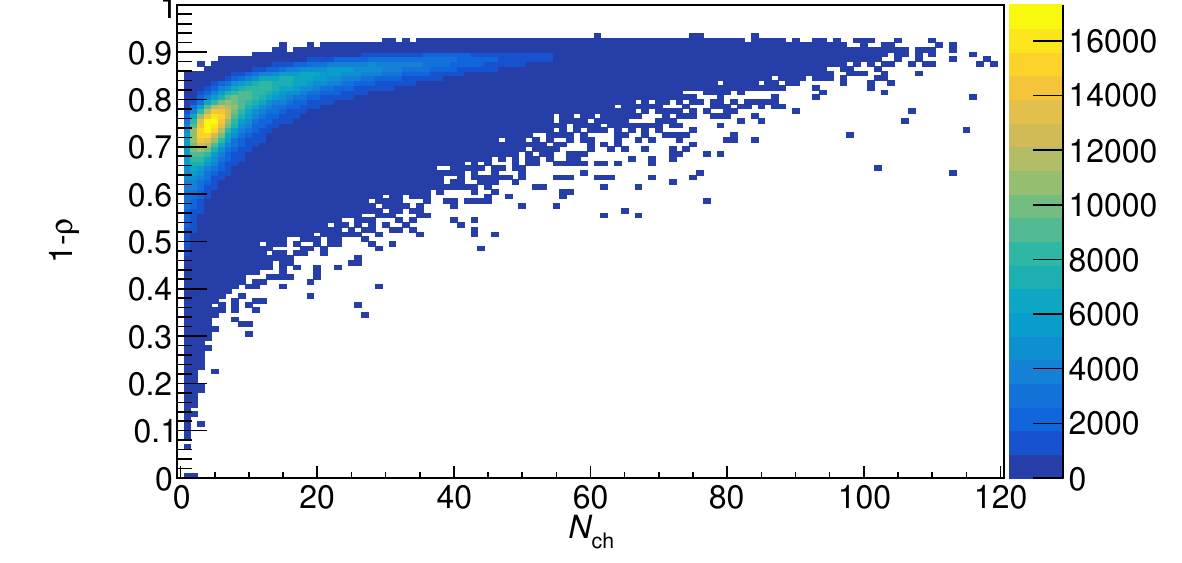}
\includegraphics[width=0.48\textwidth]{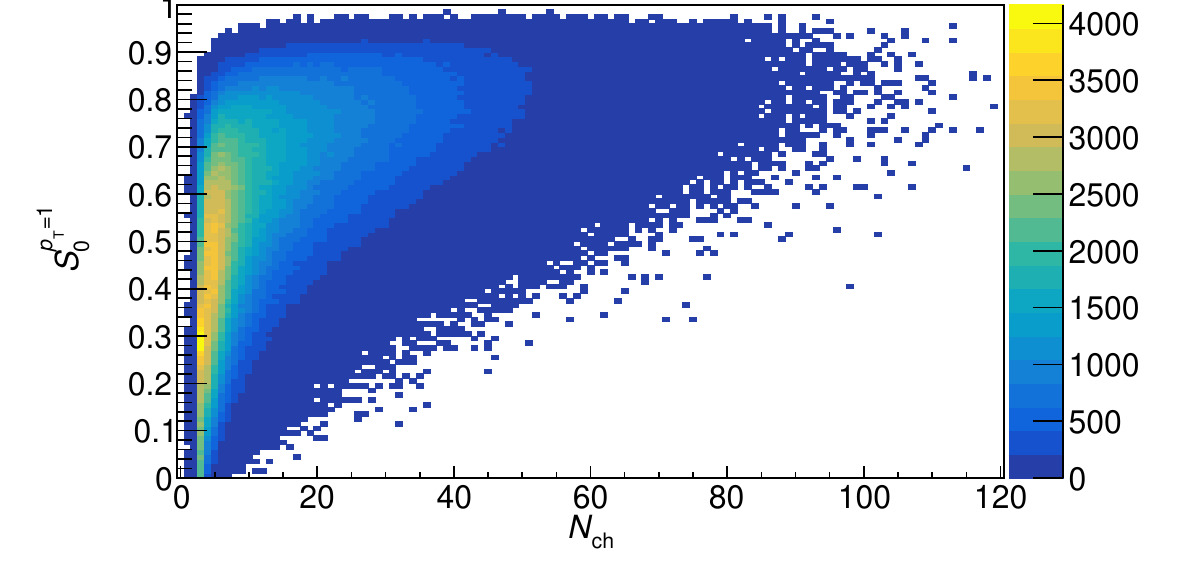}
\caption{\label{fig:correlations}Two-dimensional probability distributions of \Nch--$(1-\rho)$ (left) and \Nch--\SO (right) from PYTHIA 8 simulations at \sqs=13 TeV.}
\end{figure}
Fig.~\ref{fig:correlations} shows a relatively strong correlation between \Nch and $\rho$. For an idealized case with homogeneous variation across all cells, a clear $\rho \sim \langle \Nch \rangle^{-2}$ dependence is observed. Moreover, at lower multiplicities, $\rho$ exhibits a wide spread, while at higher multiplicities, the multi-jet topologies become typical. Although the direct dependence of \Nch on \SO is weak, a similar trend is observed as for flattenicity: at low multiplicities, \Nch spreads across the full range with lower values, but remains around unity at higher multiplicities; however, the large spread for \SO persists even at high \Nch values.

\section{The Tsallis thermometer}

The Tsallis-thermometer is a generalized two-dimensional tool that presents not only the mean energy over the number of degrees of freedom ($T$), but also the deviation from the additive, Boltzmannian case, represented by the non-extensivity parameter $q$. These parameters originate from the Tsallis entropy and exhibit correlations with system size and fluctuations~\cite{Biro:2014yoa}. Analyzing the parameters of the identified hadron spectra on the $T-q$ plane provides further insight into the strong correlations and non-extensivity of the system.

We used ALICE data for identified charged hadron production from pp collisions at $\sqs=13$ TeV, at the central pseudorapidity range $|\eta|<0.5$. The spectra of $\pi^\pm$, K$^\pm$, p and $\bar{\rm p}$ that we analyze are classified by V0M~\cite{ALICE:2020nkc}, flattenicity~\cite{ALICE:2024vaf} and spherocity~\cite{ALICE:2023bga}. Data classified by V0M is divided into 10 classes. 
Charged-particle spectra have been measured in dependence of flattenicity both for V0M-integrated data, and for the highest V0M percentile. Finally, spectra have been measured in the 10\% highest and 10\% lowest transverse spherocity percentiles in the highest mid-rapidity tracklet multiplicity percentile.

The spectra of pions, kaons, and protons were fitted with the Tsallis--Pareto distribution,
\begin{equation}
\left.\frac{\dd^{2}N}{2\pi \pT \dd\pT \dd y}\right|_{y\approx 0}
= A  \, \mT \, \left[1 + \frac{q-1}{T}(\mT-m)\right]^{-\frac{q}{q-1}} ,
\end{equation}
where $y$ and \pT are the rapidity and transverse momentum, $A$ is a normalization parameter, $T$ and $q$ are the Tsallis temperature and non-extensivity parameters respectively, $m$ is the mass of the given particle species, and $\mT = \sqrt{\pT^2 + m^2}$ is the transverse mass.
The Tsallis parameters $T$ and $q$ were extracted with the same procedure as described in our previous works~\cite{Biro:2020kve, Gyulai:2024dkq, Gyulai:2024qov}. 
A detailed description of the classification, together with the fitted Tsallis parameters for the three particle species, is in Appendix~\ref{sec:tables}.

\begin{figure}[h!]
\centerline{\includegraphics[width=0.5\textwidth]{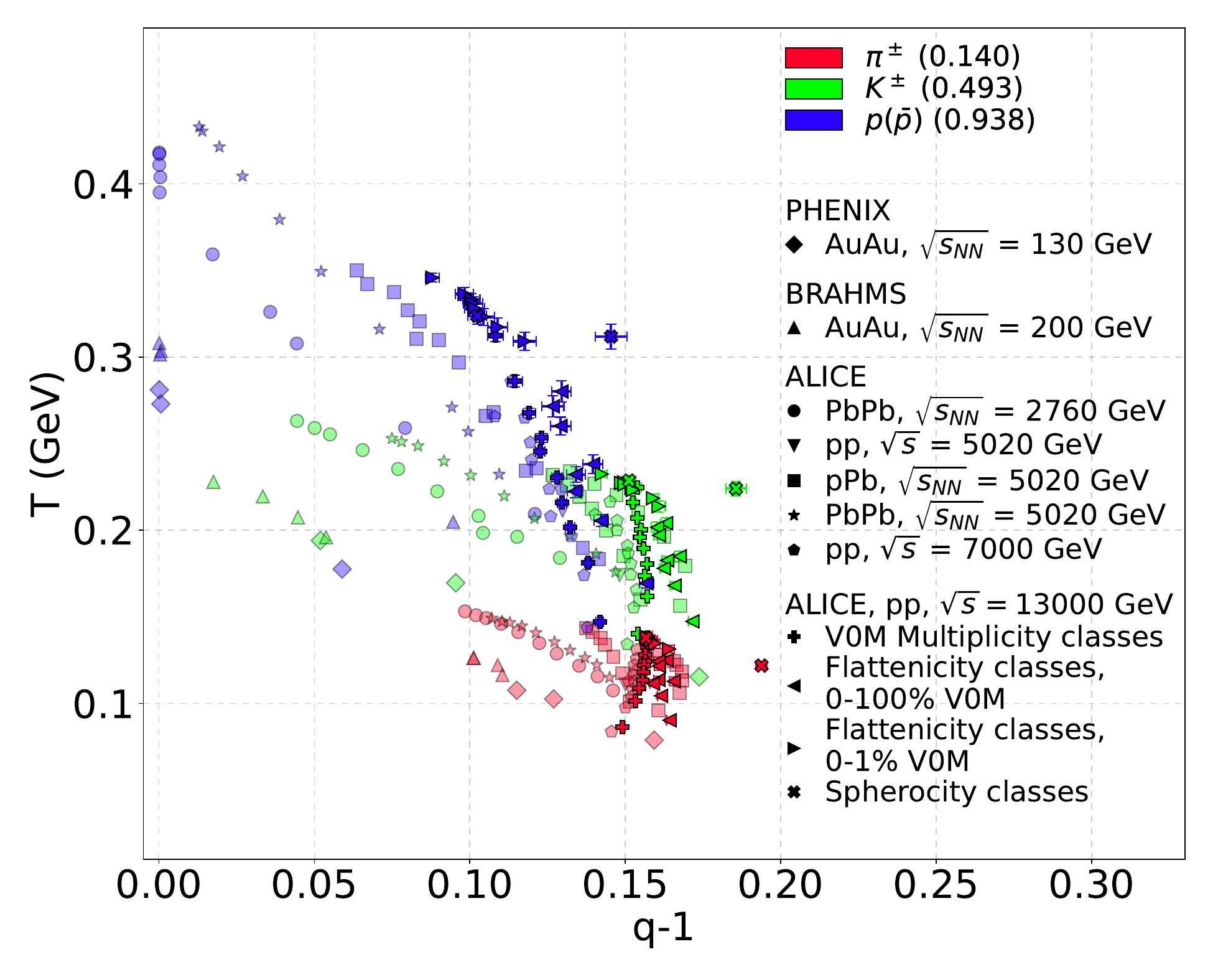}}
\caption{\label{fig:TQPiKP}The Tsallis-thermometer for $\pi^\pm$, K$^\pm$, p and $\bar{\rm p}$ for ALICE pp collisions at $\sqs=13$ TeV depending on V0M, flattenicity and spherocity~\cite{ALICE:2020nkc,ALICE:2024vaf,ALICE:2023bga} (opaque markers). The points are compared to data taken in Au--Au, pp, p--Pb and Pb--Pb collisions at lower center-of-mass energies~\cite{Biro:2020kve} (semi-transparent markers).}
\end{figure}

The $T-q$ parameter pairs are shown in the Tsallis-thermometer in Fig.~\ref{fig:TQPiKP}, together with the RHIC Au--Au as well as LHC pp, p--Pb and Pb--Pb collisions, from~\cite{Biro:2020kve}. The trends of multiplicity-dependent datasets are preserved for the presently studied $\sqs=13$ TeV data. The points corresponding to $\sqs=13$ TeV V0M  classes are located close to the $\sqs=7$ TeV points on the Tsallis-thermometer; however, they are slightly shifted towards higher $q$ and $T$ values, which is expected with the increase of energy. Points extracted from the flattenicity-classified spectra behave similarly to the multiplicity-dependent ones: a strong multiplicity-dependence within the Tsallis-thermometer can be observed, together with the mass scaling. The spherocity-dependent points, on the other hand, exhibit a different behavior compared to all other datasets, as a strong $q$ dependence can be observed. This corroborates previous observations that in events where a larger part of transverse momentum is concentrated close to the jet axis, the $q$ value is higher~\cite{Mishra:2021hnr}.

\section{Event-shape-related effects}

\begin{figure}[h!]
\includegraphics[width=0.48\textwidth]{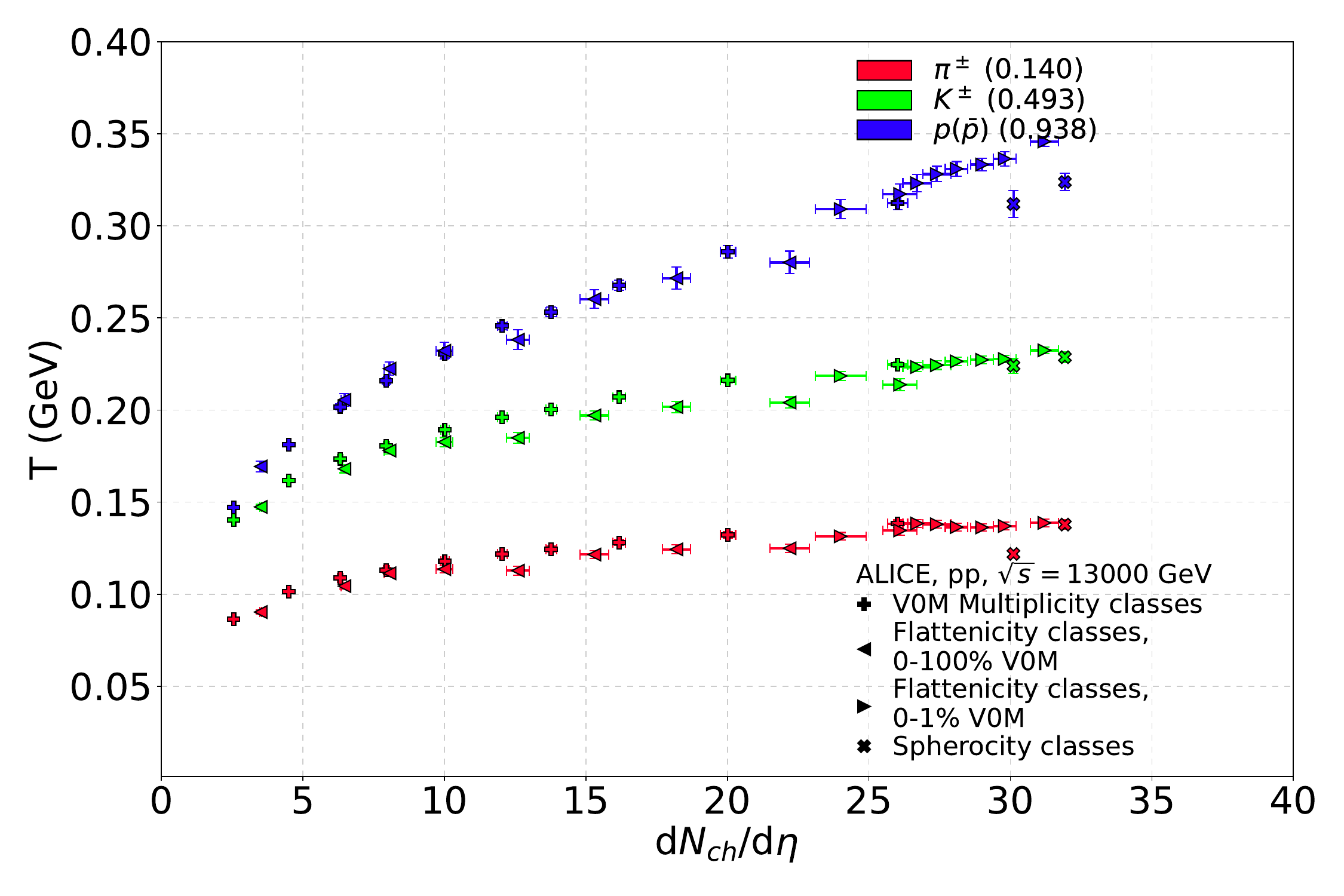}
\includegraphics[width=0.48\textwidth]{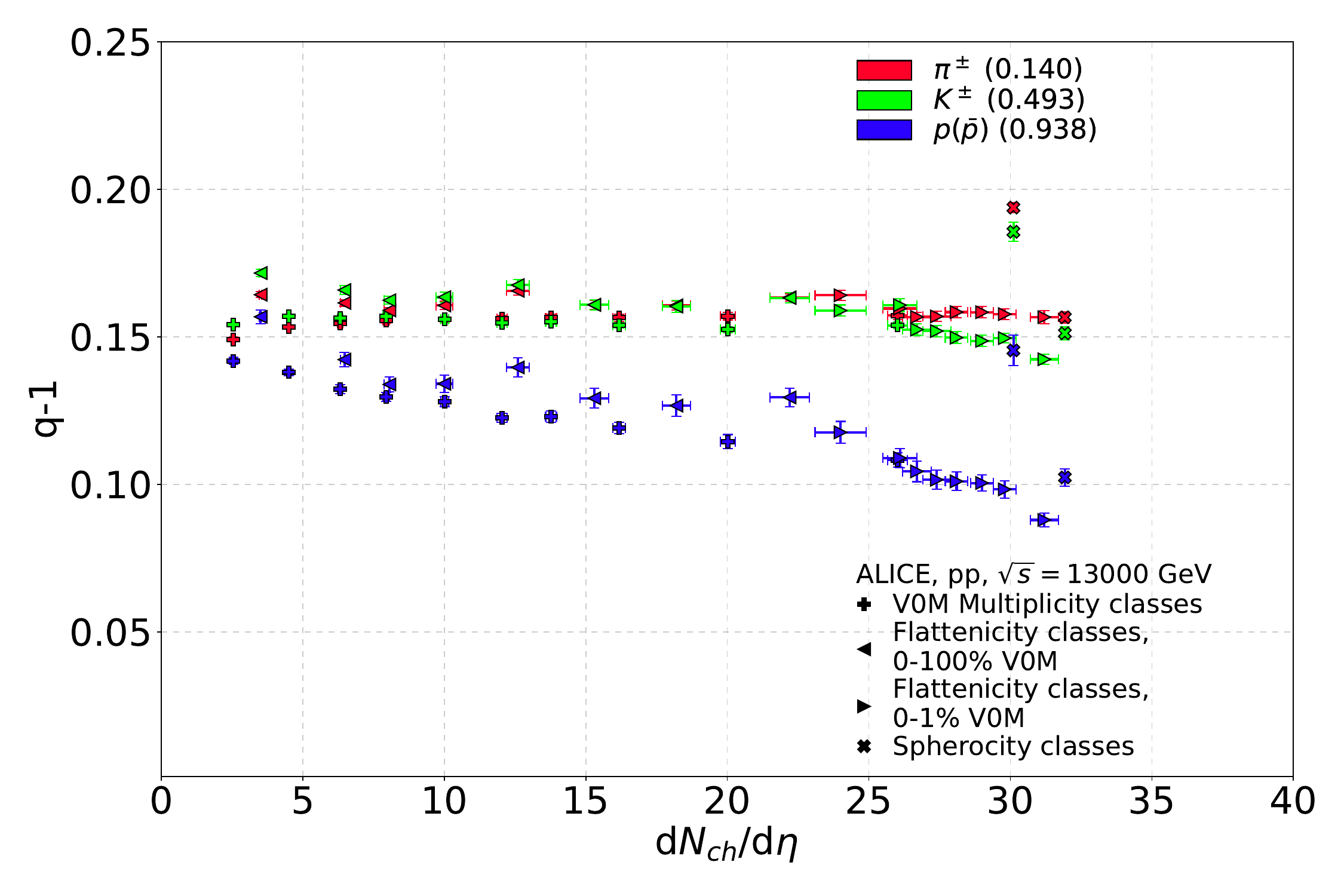}
\caption{\label{fig:allinmult}Tsallis temperature (top) and non-extensivity (bottom) parameters as a function of \dNch based on V0M, flattenicity and spherocity-dependent data~\cite{ALICE:2020nkc,ALICE:2024vaf,ALICE:2023bga}.}
\end{figure}

The Tsallis-thermometer is sensitive to event classification. To better understand the effects of multiplicity dependence of the data points, we plotted the $T$ and $q$ parameters separately, as a function of \dNch. In each data class, the average value of \dNch for each class was considered.
The dependence of the Tsallis temperature $T$ on \dNch is shown in Fig.~\ref{fig:allinmult} (top panel). For each dataset, $T$ rises monotonously with multiplicity. Multiplicity-dependent data is well complemented by flattenicity-dependent data with classes defined through V0M 0--100\%. The flattenicity-dependent points corresponding to 0-1\% V0M further continue the trend toward high multiplicity. 
Transverse spherocity, however, selects jetty events that are not part of this trend, regardless of the average V0M values, showing that the Tsallis-thermometer is sensitive of the event geometry. This interpretation is in agreement with the findings of Ref.~\cite{Mishra:2021hnr}, where Tsallis parameters of the the jet region has been shown to be distinctive from that of the regions surrounding the jets. 

Fig.~\ref{fig:allinmult} (bottom panel) shows the dependence of $q-1$ on the event multiplicity for the different datasets.
There is no strong multiplicity dependence of the $q$ parameter for pions and kaons, the data points show a flat curve. For the heavier protons, on the other hand, a slight decrease of $q$ values is observed towards higher multiplicities, as it is known from Ref.~\cite{Biro:2017arf}. The spherocity points, however, show a strikingly different behavior from the other points in case of all three particle species. While jetty and isotropic event classes correspond to almost the same average multiplicity due to the event selection, their $q$ values differ significantly, with $\Delta q \approx 0.03$--$0.05$. This indicates that the shift of the $q$ values is solely due to the differing event shapes, and not a multiplicity scaling effect, as was observed in multiplicity-dependent data. Therefore, Tsallis-thermometer is very sensitive to the event shape.
That the $q$ values are higher for more jetty events is also in agreement with the observation that $q_{\rm eq}$ in Ref.~\cite{Gyulai:2024dkq} is higher for heavy than light flavor, which is caused by the early, hard processes forming the heavy flavor spectra.
\begin{figure}[h!]
\centering
\includegraphics[width=0.48\textwidth]{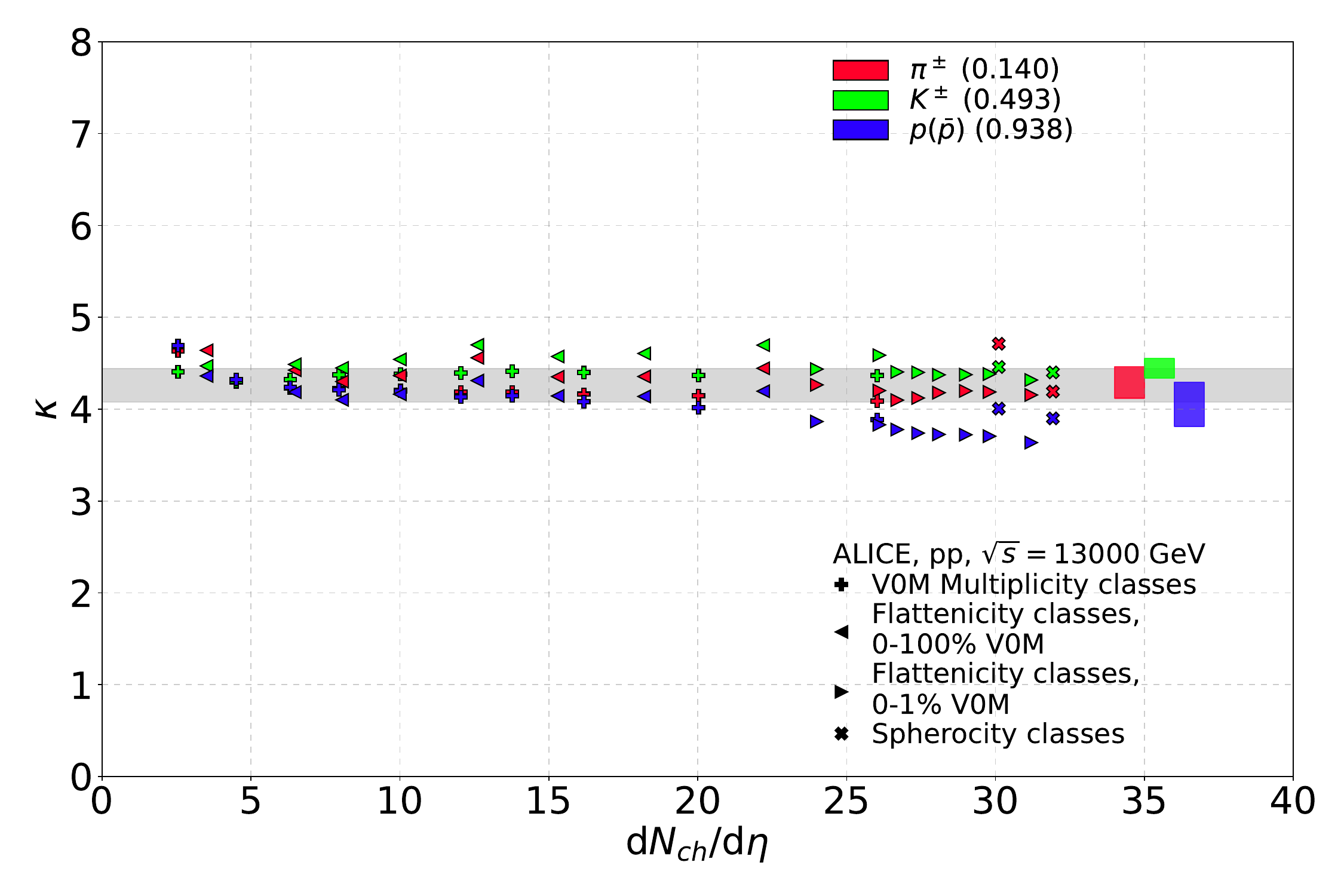}
\includegraphics[width=0.48\textwidth]{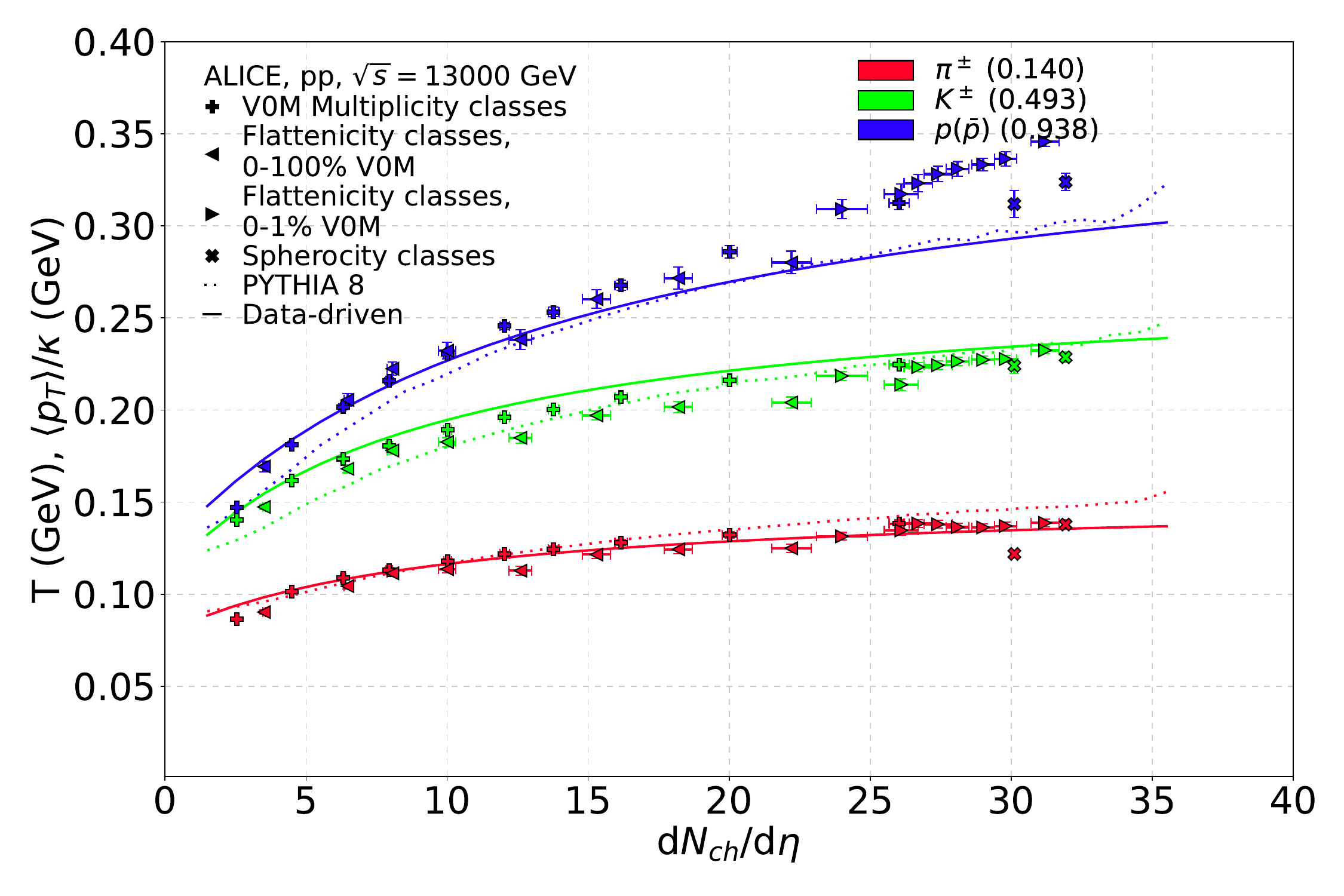}
\caption{\label{fig:kappa} Top: coefficient $\kappa$ for the various datasets (points) and the average with uncertainty (band).
The boxes represent the fit results of the individual hadron species with uncertainties.
Bottom: Tsallis temperature parameters as a function of \dNch based on V0M, flattenicity and spherocity-dependent data~\cite{ALICE:2020nkc,ALICE:2024vaf,ALICE:2023bga}, compared to $\meanpT/\kappa$ extracted with the data-driven method (solid line) and from PYTHIA~8 simulations (dotted line).
}
\end{figure}

While the Tsallis parameters are sensitive to the peculiarities of the full spectra, the mean momentum \meanpT, connected to the first moment of the spectra, also carries some of the information. It has been shown that there is a simple proportional relation between the Boltzmann temperature and \meanpT~\cite{Gardim:2019xjs,Gardim:2024zvi}. This also works with the Tsallis temperature $T$, albeit with a different proportionality parameter $\kappa$, defined as
\begin{equation}
    \kappa=\frac{T}{\meanpT} .
\end{equation}
To determine $\kappa$, we used a parametrization
\begin{equation}
\langle p_{\rm T}\rangle = a - b ( c - \langle \dNch \rangle)^{-1}      
\end{equation}
based on pion, kaon and proton data from ALICE~\cite{ALICE:2020nkc}. 
The average coefficient over all the points is $\kappa = 4.26 \pm 0.18$ from the data-driven method, as shown in Fig.~\ref{fig:kappa} (top panel). The separate averages over specific particle species are consistent with each other. Note that although this value of $\kappa$ is substantially different than the $\kappa \approx 3$ case of the Boltzmann\,--\,Gibbs statistics~\cite{Gardim:2019xjs}, the same linear scaling holds.
For comparison, we also generated pion, kaon, and proton spectra using PYTHIA (version 8.3)~\cite{Bierlich:2022pfr} with the Monash tune~\cite{Skands:2014pea}, and computed the \meanpT values with respect to the multiplicity \dNch. This leads to a somewhat different value of $\kappa = 4.71 \pm 0.20$. It is to be noted that kaon and proton $\langle p_{\rm T}\rangle$ are not described well by PYTHIA~8~\cite{ALICE:2020nkc}.
In the bottom panel of Fig.~\ref{fig:kappa}, the obtained \meanpT values divided by the correcponding average $\kappa$ value are compared to the Tsallis temperatures extracted from the various datasets. A good correspondence can be observed between the $T$ and the \meanpT values extracted by the data-driven method.

\begin{figure}[h!]
\centerline{
\includegraphics[width=0.5\textwidth]{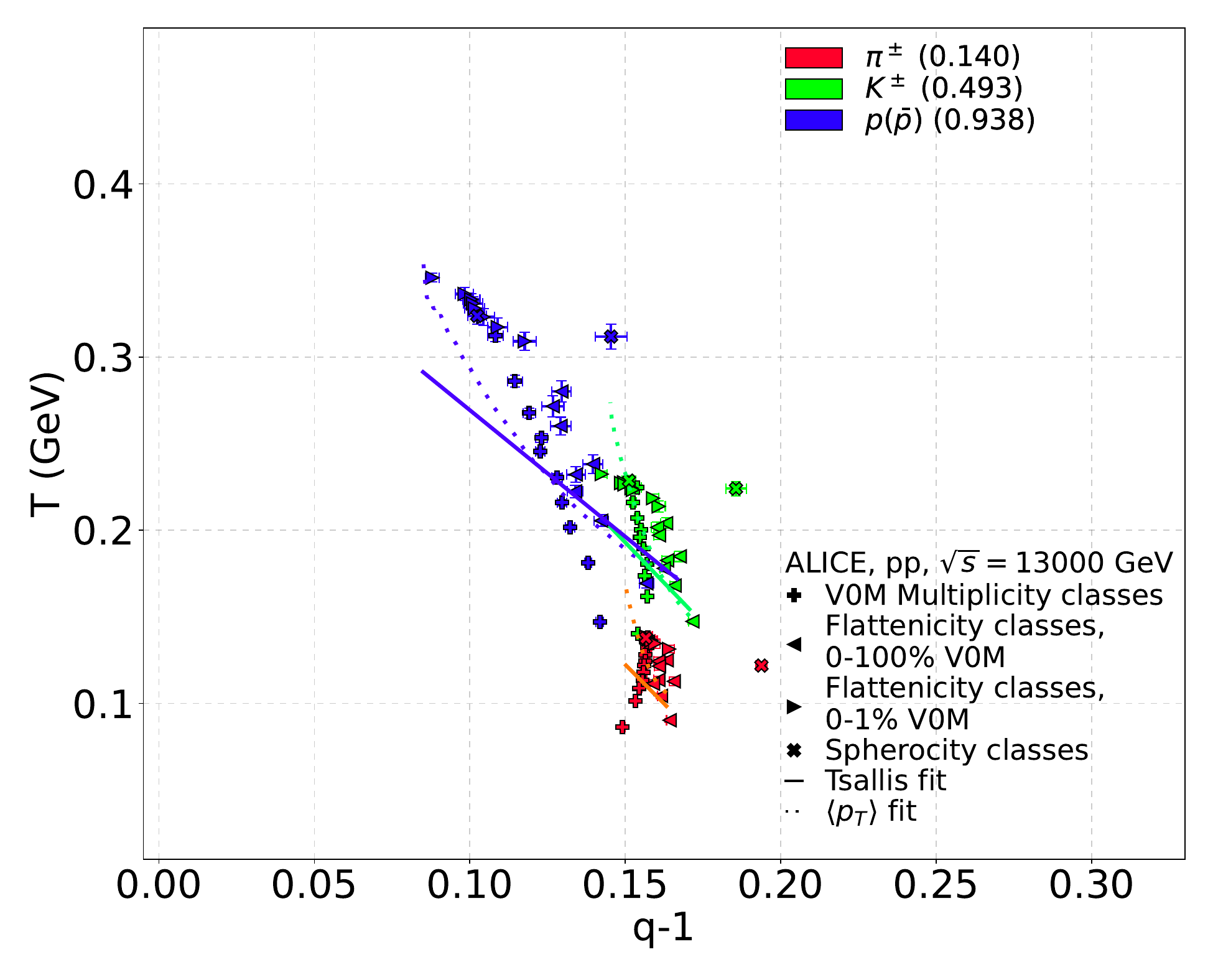}}
\caption{\label{fig:T_meanpT}%
The Tsallis-thermometer for $\pi^\pm$, K$^\pm$, p and $\bar{\rm p}$ for ALICE pp collisions at $\sqs=13$ TeV depending on V0M, flattenicity and spherocity~\cite{ALICE:2020nkc,ALICE:2024vaf,ALICE:2023bga} (opaque markers), compared to Tsallis fit and a \meanpT fit calculations based on PYTHIA 8 (solid and dotted lines, respectively).}
\end{figure}

In Fig.~\ref{fig:T_meanpT} the Tsallis-thermometer is shown again, concentrating on the current $\sqs=13$ TeV data. The solid lines show the Tsallis fit as detailed in Ref.~\cite{Biro:2020kve}. The dotted lines have the same calculation for the $q$ parameter, however, the $T$ parameter is instead calculated from the generated \meanpT values as $\meanpT/\kappa$. 
Both fits generally recreate the position of data points on the Tsallis-thermometer. 
While multiplicity and flattenicity classes exhibit a linear trend among all investigated hadron species, the spherocity-dependent classifications show a distinct behavior, with a deviating $\kappa$ for protons. This highlights the sensitivity of the Tsallis-thermometer to event shape and geometry, providing a unique way for event selection.

To understand how event classification affects the points on the Tsallis-thermometer, it is necessary to examine the two classification methods using transverse spherocity and flattenicity. While flattenicity correlates weakly with multiplicity in the regime of $1-\rho \in [0.5:0.9]$, this weak dependence has no direct effect on the tail of the spectra. (The spectrum ratios with different rapidities, $Q_{\rm pp}$, are flat toward high \pT~\cite{ALICE:2024vaf}.) 
In contrast, the spherocity-dependent transverse momentum distribution changes due to the redistribution of contributions from certain multiplicity classes (see Fig.\ 4 in Ref.~\cite{ALICE:2023bga}).
While the body of the distribution, represented by the temperature parameter, remains relatively unchanged, there is a strong variation in the power $n=-q/(1-q)$ of the tail of the $p_T$ distribution, reflecting a greater contribution from jetty and hard events toward smaller spherocity values. The variation observed in Ref.~\cite{Mishra:2021hnr} is $\Delta n \approx 0.05$, and other works~\cite{ALICE:2023bga,ALICE:2020nkc,Mishra:2018pio} show similar trends. This is in agreement with the observed $\Delta q \approx 0.03$--$0.05$. Applying the spherocity-classified data at the highest 10 percentiles, more jetty (hard) events are selected with higher non-extensivity parameter values, resulting in a horizontal shift of the corresponding data points on the Tsallis-thermometer.

\section{Conclusions}

We have studied identified hadron spectra in pp collisions at $\sqs = 13$ TeV within the framework of Tsallis statistics, using ALICE data classified by multiplicity (V0M), flattenicity, and transverse spherocity. The extracted Tsallis parameters were placed on the Tsallis-thermometer and compared to previous results from lower energies and from larger collision systems.

We find that both multiplicity and flattenicity classes preserve the expected scaling trends of the Tsallis temperature and non-extensivity parameter, with $T$ rising monotonically with charged-particle density and $q$ showing only a weak dependence. In contrast, spherocity-dependent classifications reveal a qualitatively different behavior, with considerably higher $q$ values in jetty events. This highlights the sensitivity of the Tsallis framework to the event geometry and the underlying production mechanisms. The non-extensivity parameter can thus be linked to the role of early hard processes, consistent with previous findings for heavy-flavor spectra. 

Our findings are also in alignment with the detailed analysis of data from small to large collisional systems: pp, pA, and AA~\cite{Biro:2020kve}. A future non-extensive analysis of collisions of light O--O and Ne--Ne systems recently recorded at the Large Hadron Collider may provide new insights to the thermodynamical properties around the onset of QGP signatures.

We confirmed a proportional relation between the Tsallis temperature and the mean transverse momentum, quantified by $\kappa = 4.26 \pm 0.18$, using a data-driven parametrization. This scaling provides a complementary approach to estimate the Tsallis temperature from the first moment of the spectra, and it successfully reproduces the position of the experimental points on the Tsallis-thermometer.

In summary, the present study demonstrates that the Tsallis-thermometer is a powerful tool for disentangling the interplay between multiplicity, event shape, and hadron species in high-energy collisions. Applications of this framework to larger systems in the future  may help the deeper understanding of the connection between non-extensive thermodynamics, collective behavior, and the microscopic processes that shape particle spectra.

\section*{Acknowledgements}
This work has been supported by the Hungarian National Research, Development and Innovation Office (NKFIH) under the contract numbers NKFIH 
NKKP ADVANCED\_25-153456, NEMZ\_KI-2022-00058, 2025-1.1.5-NEMZ\_KI-2025-00002, 2025-1.1.5-NEMZ\_KI-2025-00005, 2025-1.1.5-NEMZ\_KI-2025-00013 and 2024-1.2.5-TET-2024-00022, the FuSe COST Action CA-24101 and the Wigner Scientific Computing Laboratory (WSCLAB).
We are also grateful for the possibility to use HUN-REN Cloud (see Héder et al. 2022; https://science-cloud.hu/) which helped us achieve the results published in this paper.

\appendix

\section{Event classification and Tsallis fit parameters}
\label{sec:tables}

Data from Ref.~\cite{ALICE:2020nkc} contains spectra of $\pi^\pm$, K$^\pm$, p and $\bar{\rm p}$ in 10 different V0M classes. Table~\ref{tab:multiplicity} shows classification by V0M together with the fitted Tsallis parameters $T$ and $q$. A strong correlation between the average charged-hadron multiplicity and V0M is observable.
\begin{table}[p]
    \caption{\label{tab:multiplicity}V0M classes, from Ref.~\cite{ALICE:2020nkc}.}
    %\resizebox{\columnwidth}{!}{
    \begin{tabular}{cccccc}
    \toprule
    V0M & $\sigma/\sigma_{{\rm INEL}>0}$  & $\left<\dd \Nch/\dd\eta\right>$ & Hadron & T (GeV) & q\\
    class & (\%) &&type&&\\
    \midrule
     \multirow{3}{*}{I} & \multirow{3}{*}{0--0.92} & \multirow{3}{*}{$26.02\pm0.35$} & $\pi^\pm$& $0.14\pm0.00$& $1.16\pm0.00$\\
     &&& K$^\pm$ & $0.22\pm0.00$& $1.15\pm0.00$\\
     &&& p$(\overline {\rm p})$ & $0.31\pm0.00$& $1.11\pm0.00$\\
             \midrule
    \multirow{3}{*}{II} & \multirow{3}{*}{0.92--4.6} & \multirow{3}{*}{$20.02\pm0.27$} & $\pi^\pm$& $0.13\pm0.00$& $1.16\pm0.00$ \\
    &&& K$^\pm$ & $0.22\pm0.00$& $1.15\pm0.00$ \\
    &&& p$(\overline {\rm p})$ & $0.29\pm0.00$& $1.11\pm0.00$\\
            \midrule
    \multirow{3}{*}{III} & \multirow{3}{*}{4.6--9.2} & \multirow{3}{*}{$16.17\pm0.22$}  &$\pi^\pm$& $0.13\pm0.00$& $1.16\pm0.00$\\
    &&& K$^\pm$ & $0.21\pm0.00$& $1.15\pm0.00$ \\
    &&& p$(\overline {\rm p})$ & $0.27\pm0.00$& $1.12\pm0.00$ \\
            \midrule
    \multirow{3}{*}{IV} & \multirow{3}{*}{9.2--13.8} & \multirow{3}{*}{$13.77\pm0.19$} & $\pi^\pm$& $0.12\pm0.00$& $1.16\pm0.00$  \\
    &&& K$^\pm$ & $0.20\pm0.00$& $1.16\pm0.00$ \\
    &&& p$(\overline {\rm p})$ & $0.25\pm0.00$& $1.12\pm0.00$\\
            \midrule
    \multirow{3}{*}{V} & \multirow{3}{*}{13.8--18.4} & \multirow{3}{*}{$12.04\pm0.17$} & $\pi^\pm$& $0.12\pm0.00$& $1.16\pm0.00$  \\
    &&& K$^\pm$ & $0.20\pm0.00$& $1.15\pm0.00$  \\
    &&& p$(\overline {\rm p})$ & $0.25\pm0.00$& $1.12\pm0.00$ \\
            \midrule
    \multirow{3}{*}{VI} & \multirow{3}{*}{18.4--27.6} & \multirow{3}{*}{$10.02\pm0.14$} & $\pi^\pm$& $0.12\pm0.00$& $1.16\pm0.00$ \\
    &&& K$^\pm$ & $0.19\pm0.00$& $1.16\pm0.00$  \\
    &&& p$(\overline {\rm p})$ & $0.23\pm0.00$& $1.13\pm0.00$ \\
            \midrule
    \multirow{3}{*}{VII} & \multirow{3}{*}{27.6--36.8} & \multirow{3}{*}{$7.95\pm0.11$} & $\pi^\pm$& $0.11\pm0.00$& $1.16\pm0.00$  \\
    &&& K$^\pm$ & $0.18\pm0.00$& $1.16\pm0.00$ \\
    &&& p$(\overline {\rm p})$ & $0.22\pm0.00$& $1.13\pm0.00$  \\
            \midrule
    \multirow{3}{*}{VIII} & \multirow{3}{*}{36.8--46.0} & \multirow{3}{*}{$6.32\pm0.09$}& $\pi^\pm$& $0.11\pm0.00$& $1.15\pm0.00$  \\
    &&& K$^\pm$ & $0.17\pm0.00$& $1.16\pm0.00$  \\
    &&& p$(\overline {\rm p})$ & $0.20\pm0.00$& $1.13\pm0.00$  \\
            \midrule
    \multirow{3}{*}{IX} & \multirow{3}{*}{46.0--64.5} & \multirow{3}{*}{$4.50\pm0.07$} & $\pi^\pm$& $0.10\pm0.00$& $1.15\pm0.00$  \\
    &&& K$^\pm$ & $0.16\pm0.00$& $1.16\pm0.00$  \\
    &&& p$(\overline {\rm p})$ & $0.18\pm0.00$& $1.14\pm0.00$ \\
            \midrule
    \multirow{3}{*}{X} & \multirow{3}{*}{64.5--100} & \multirow{3}{*}{$2.55\pm0.04$}  & $\pi^\pm$& $0.09\pm0.00$& $1.15\pm0.00$  \\
    &&& K$^\pm$ & $0.14\pm0.00$& $1.15\pm0.00$  \\
    &&& p$(\overline {\rm p})$ & $0.15\pm0.00$& $1.14\pm0.00$ \\
     \bottomrule
    \end{tabular}
   %}
\end{table}
Charged-particle spectra have been measured in dependence of flattenicity both for V0M-integrated data, and for the highest V0M percentile~\cite{ALICE:2024vaf}. This classification is summarized in Table~\ref{tab:flattenicity}, together with the fitted Tsallis parameters $T$ and $q$.
While in the V0M-integrated case, a strong correlation between the flattenicity and the average charged-hadron multiplicity is present, there is a much weaker correlation in the high-V0M classes.
\begin{table}[p]
\caption{Flattenicity classes, from Ref.~\cite{ALICE:2024vaf}.}
    \label{tab:flattenicity}
    %\resizebox{\columnwidth}{!}{
    
    \begin{tabular}{cccccc}
    \toprule
    \multicolumn{6}{c}{V0M percentile 0--100\%} \\
    $1-\rho$ & $1-\rho$ (\%) & \multirow{2}{*}{$\left<\dd \Nch/\dd\eta\right>$} & Hadron & \multirow{2}{*}{T (GeV)} & \multirow{2}{*}{q} \\
    class&percentile&&type&&\\
    \midrule
          \multirow{3}{*}{I} & \multirow{3}{*}{0--1} & \multirow{3}{*}{$22.2\pm0.7$} & $\pi^\pm$& $0.12\pm0.00$& $1.16\pm0.00$\\
        &&& K$^\pm$ & $0.20\pm0.00$& $1.16\pm0.00$ \\
        &&& p$(\overline {\rm p})$ & $0.28\pm0.00$& $1.13\pm0.01$ \\
        \midrule
          \multirow{3}{*}{II} & \multirow{3}{*}{1--5} & \multirow{3}{*}{$18.2\pm0.5$} &$\pi^\pm$&$0.12\pm0.00$& $1.16\pm0.00$ \\
        &&& K$^\pm$ & $0.20\pm0.00$& $1.16\pm0.00$ \\
        &&& p$(\overline {\rm p})$ & $0.27\pm0.00$& $1.13\pm0.01$ \\
        \midrule
          \multirow{3}{*}{III} & \multirow{3}{*}{5--10} & \multirow{3}{*}{$15.3\pm0.5$} &$\pi^\pm$&$0.12\pm0.00$& $1.16\pm0.00$ \\
        &&& K$^\pm$ & $0.20\pm0.00$& $1.16\pm0.00$ \\
        &&& p$(\overline {\rm p})$ & $0.26\pm0.00$& $1.13\pm0.01$ \\
        \midrule
          \multirow{3}{*}{IV} & \multirow{3}{*}{10--20} & \multirow{3}{*}{$12.6\pm0.4$} &$\pi^\pm$&$0.11\pm0.00$& $1.17\pm0.00$ \\
        &&& K$^\pm$ & $0.18\pm0.00$& $1.17\pm0.00$ \\
        &&& p$(\overline {\rm p})$ & $0.24\pm0.00$& $1.14\pm0.01$ \\
        \midrule
    \multirow{3}{*}{V} & \multirow{3}{*}{20--30} & \multirow{3}{*}{$10.0\pm0.3$}&$\pi^\pm$&$0.11\pm0.00$& $1.16\pm0.00$ \\
        &&& K$^\pm$ & $0.18\pm0.00$& $1.16\pm0.00$ \\
        &&& p$(\overline {\rm p})$ & $0.23\pm0.00$& $1.13\pm0.00$ \\
        \midrule
    \multirow{3}{*}{VI} & \multirow{3}{*}{30--40} & \multirow{3}{*}{$8.06\pm0.19$}&$\pi^\pm$&$0.11\pm0.00$& $1.16\pm0.00$ \\
        &&& K$^\pm$ & $0.18\pm0.00$& $1.16\pm0.00$ \\
        &&& p$(\overline {\rm p})$ & $0.22\pm0.00$& $1.13\pm0.00$ \\
        \midrule
    \multirow{3}{*}{VII} & \multirow{3}{*}{40--50} & \multirow{3}{*}{$6.47\pm0.13$}&$\pi^\pm$&$0.10\pm0.00$& $1.16\pm0.00$ \\
        &&& K$^\pm$ & $0.17\pm0.00$& $1.17\pm0.00$ \\
        &&& p$(\overline {\rm p})$ & $0.21\pm0.00$& $1.14\pm0.00$ \\
        \midrule
    \multirow{3}{*}{VIII} & \multirow{3}{*}{50--100} & \multirow{3}{*}{$3.51\pm0.04$}&$\pi^\pm$&$0.09\pm0.00$& $1.16\pm0.00$ \\
        &&& K$^\pm$ & $0.15\pm0.00$& $1.17\pm0.00$ \\
        &&& p$(\overline {\rm p})$ & $0.17\pm0.00$& $1.16\pm0.00$ \\
    \bottomrule\toprule
    \multicolumn{6}{c}{V0M percentile 0--1\%} \\
    $1-\rho$ & $1-\rho$ (\%) & \multirow{2}{*}{$\left<\dd \Nch/\dd\eta\right>$} & Hadron & \multirow{2}{*}{T (GeV)} & \multirow{2}{*}{q} \\
    class&percentile&&type&&\\
    \midrule
    \multirow{3}{*}{I} & \multirow{3}{*}{0--1} & \multirow{3}{*}{$31.2\pm0.5$} & $\pi^\pm$&$0.14\pm0.00$& $1.16\pm0.00$\\
        &&& K$^\pm$ & $0.23\pm0.00$& $1.14\pm0.00$ \\
        &&& p$(\overline {\rm p})$ & $0.35\pm0.00$& $1.09\pm0.00$ \\
        \midrule
          \multirow{3}{*}{II} & \multirow{3}{*}{1--5} & \multirow{3}{*}{$29.8\pm0.4$} & $\pi^\pm$&$0.14\pm0.00$& $1.16\pm0.00$ \\
        &&& K$^\pm$ & $0.23\pm0.00$& $1.15\pm0.00$ \\
        &&& p$(\overline {\rm p})$ & $0.34\pm0.00$& $1.10\pm0.00$ \\
        \midrule
          \multirow{3}{*}{III} & \multirow{3}{*}{5--10} & \multirow{3}{*}{$29.0\pm0.4$} & $\pi^\pm$&$0.14\pm0.00$& $1.16\pm0.00$ \\
        &&& K$^\pm$ & $0.23\pm0.00$& $1.15\pm0.00$ \\
        &&& p$(\overline {\rm p})$ & $0.33\pm0.00$& $1.10\pm0.00$ \\
        \midrule
          \multirow{3}{*}{IV} & \multirow{3}{*}{10--20} & \multirow{3}{*}{$28.1\pm0.4$} & $\pi^\pm$&$0.14\pm0.00$& $1.16\pm0.00$ \\
        &&& K$^\pm$ & $0.23\pm0.00$& $1.15\pm0.00$ \\
        &&& p$(\overline {\rm p})$ & $0.33\pm0.00$& $1.10\pm0.00$ \\
        \midrule
    \multirow{3}{*}{V} & \multirow{3}{*}{20--30} & \multirow{3}{*}{$27.4\pm0.5$} & $\pi^\pm$&$0.14\pm0.00$& $1.16\pm0.00$ \\
        &&& K$^\pm$ & $0.22\pm0.00$& $1.15\pm0.00$ \\
        &&& p$(\overline {\rm p})$ & $0.33\pm0.00$& $1.10\pm0.00$ \\
        \midrule
    \multirow{3}{*}{VI} & \multirow{3}{*}{30--40} & \multirow{3}{*}{$26.7\pm0.5$} & $\pi^\pm$&$0.14\pm0.00$& $1.16\pm0.00$ \\
        &&& K$^\pm$ & $0.22\pm0.00$& $1.15\pm0.00$ \\
        &&& p$(\overline {\rm p})$ & $0.32\pm0.00$& $1.10\pm0.00$ \\
        \midrule
    \multirow{3}{*}{VII} & \multirow{3}{*}{40--50} & \multirow{3}{*}{$26.1\pm0.6$} & $\pi^\pm$&$0.13\pm0.00$& $1.16\pm0.00$ \\
        &&& K$^\pm$ & $0.21\pm0.00$& $1.16\pm0.00$ \\
        &&& p$(\overline {\rm p})$ & $0.32\pm0.01$& $1.11\pm0.00$ \\
        \midrule
    \multirow{3}{*}{VIII} & \multirow{3}{*}{50--100} & \multirow{3}{*}{$24.0\pm0.9$} & $\pi^\pm$&$0.13\pm0.00$& $1.16\pm0.00$ \\
        &&& K$^\pm$ & $0.22\pm0.00$& $1.16\pm0.00$ \\
        &&& p$(\overline {\rm p})$ & $0.31\pm0.01$& $1.12\pm0.00$ \\
    \bottomrule
    \end{tabular}

    %}
\end{table}
Finally, Ref.~\cite{ALICE:2023bga} contains spectra for the 10\% highest and 10\% lowest transverse spherocity percentiles as shown in Table~\ref{tab:spherocity} together with the fitted Tsallis parameters $T$ and $q$. The average charged-hadron multiplicity depends on the spherocity class only very weakly.
\begin{table}[p]

    \caption{Transverse spherocity classification, following Ref.~\cite{ALICE:2023bga}}
    \label{tab:spherocity}
    %\resizebox{\columnwidth}{!}{

    \begin{tabular}{cccccc}
        \toprule
        \multicolumn{6}{c}{$N_{\rm tracklet}$ percentile 0--1\%} \\
        $\SO$ & $\SO$ (\%)  &  \multirow{2}{*}{$\left<\dd \Nch/\dd\eta\right>$} & Hadron & \multirow{2}{*}{T (GeV)} & \multirow{2}{*}{q}\\
        class & percentile && type & & \\
        \midrule
        \multirow{3}{*}{Jet-like} & \multirow{3}{*}{0--10} & \multirow{3}{*}{$30.11\pm0.06$} & $\pi^{\pm}$ & $0.12\pm0.00$& $1.19\pm0.00$  \\
        &&& K$^\pm$ & $0.22\pm0.00$& $1.19\pm0.00$ \\
        &&& p$(\overline {\rm p})$ & $0.31\pm0.01$& $1.15\pm0.01$ \\
        \midrule
        \multirow{3}{*}{Isotropic} & \multirow{3}{*}{90--100} & \multirow{3}{*}{$31.93\pm0.06$} & $\pi^{\pm}$ & $0.14\pm0.00$& $1.16\pm0.00$  \\
        &&& K$^\pm$ & $0.23\pm0.00$& $1.15\pm0.00$ \\
        &&& p$(\overline {\rm p})$ & $0.32\pm0.00$& $1.10\pm0.00$ \\
        \bottomrule
    \end{tabular}
    
    %}
\end{table}
For comparison, in the tables we indicated the average \Nch per pseudorapidity unit for each event class.

\bibliography{TsallisEvtshape}

\end{document}